\documentclass[a4paper,11pt]{elsarticle}

\usepackage[margin=2.5cm]{geometry}
\usepackage[utf8x]{inputenc}
\usepackage{epsfig}
\usepackage{graphicx}
\usepackage{amsmath}
\usepackage{amssymb}
\usepackage{hyperref}

\newcommand{\SPC}{\hphantom{.}}

\begin{document}

  \begin{frontmatter}

    \title{Properties of strange vector mesons in dense and hot matter}
    
    \author{Andrej Ilner$^1$, Daniel Cabrera$^{1,2}$, Pornrad Srisawad$^3$, Elena Bratkovskaya$^{1,2}$}

    \address{$^1$ Institut f\"ur Theoretische Physik, Johann Wolfgang Goethe-Universit\"at Frankfurt am Main, 60438 Frankfurt am Main, Germany}
    \address{$^2$ Frankfurt Institute for Advanced Studies (FIAS), 60438 Frankfurt am Main, Germany}
    \address{$^3$ Naresuan University, Faculty of Science, Phitsanulok 65000, Thailand}



    \begin{abstract}
We investigate the in-medium properties of strange vector mesons ($K^*$ and $\bar K^*$) in dense and hot nuclear matter based on chirally motivated models of the meson selfenergies. We parameterise medium effects as density or temperature dependent effective masses and widths, obtain the vector meson spectral functions within a Breit-Wigner prescription (as often used in transport simulations) and study whether such an approach can retain the essential features of full microscopic calculations. For $\mu_B\ne 0$ the medium corrections arise from $\bar K^* (K^*) N$ scattering and the $\bar K^* (K^*) \to \bar K (K) \pi$ decay mode (accounting for in-medium $\bar K (K)$ dynamics).
We calculate the scattering contribution to the $K^*$ selfenergy based on the hidden local symmetry formalism for vector meson nucleon interactions, whereas for the $\bar K^*$ selfenergy we implement recent results from a selfconsistent coupled-channel determination within the same approach. For $\mu_B\simeq 0$ and finite temperature we rely on a phenomenological approach for the kaon selfenergy in a hot pionic medium consistent with chiral symmetry, and evaluate the $\bar K^* (K^*) \to \bar K (K) \pi$ decay width. The emergence of a mass shift at finite temperature is studied with a dispersion relation over the imaginary part of the vector meson selfenergy.
    \end{abstract}

  \end{frontmatter}

  \section{Introduction}
The features of strongly interacting matter in a broad range of temperatures and densities has been a subject of great interest in the last decades, in connection with fundamental aspects of the strong interaction such as the nature of deconfinement or the physical mechanism of chiral symmetry restoration \cite{Rapp:1999ej}. Hadrons with strangeness, in particular, have been matter of intense investigation regarding the study of exotic atoms \cite{Friedman:2007zza}, the analysis of strangeness production in heavy-ion collisions (HICs) \cite{Fuchs:2005zg,Forster:2007qk,Hartnack:2011cn} and the microscopic dynamics ruling the composition of neutron stars \cite{Kaplan:1986yq}.

Understanding the dynamics of light strange mesons in a nuclear environment has posed a challenge to theoretical models, amongst different reasons, due to the fact that the low-density approximation
$2\omega V_{\rm opt} \simeq t \rho_N$ is unable to
to describe the interaction of the $\bar K$ meson with the medium, as it is concluded from the phenomenology of kaonic atoms \cite{Friedman:2007zza}: whereas the $\bar K N$ scattering amplitude ($T$-matrix) is repulsive at low energies, the data require an attractive optical potential.
The early departure of the $\bar K$ nuclear potential from the low-density result is tied to the presence, below the $\bar K N$ threshold, of the $\Lambda(1405)$ resonance.
The onset of an attractive $\bar K N$ interaction at low densities is a consequence of an upper shift of the excitation energy of this resonance induced by Pauli blocking on the intermediate nucleon states \cite{Koch:1994mj,Waas:1996xh,Waas:1996fy,Lutz:1997wt}. The latter changes the real part of the $\bar K N$ scattering amplitude from repulsive in free space to attractive in a nuclear medium already at very low densities, whereas additional medium effects such as the selfenergy of mesons in related coupled channels and the binding of baryons in the nuclear environment bring the $\Lambda (1405)$ back to (almost) its vacuum position.
This peculiar behaviour of the $\bar K$ meson interaction with the nuclear medium is successfully described in chirally motivated coupled-channel approaches implementing the exact unitarity of the scattering matrix 
\cite{Waas:1996xh,Waas:1996fy,Lutz:1997wt,Weise:2008aj}
and a selfconsistent evaluation of the kaon selfenergy \cite{Lutz:1997wt,Ramos:1999ku,Tolos:2000fj,Lutz:2007bh}.
An attractive potential of about 40-60~MeV at normal nuclear matter density is obtained when selfconsistency is implemented in chiral approaches (approximately half the size when only Pauli blocking is implemented), rather shallow as compared to relativistic mean-field calculations \cite{Schaffner:1996kv} or required by phenomenological analysis of kaonic atom data \cite{Friedman:2007zza} with density dependent potentials including non-linearities. Yet, this potential is able to reproduce the data from kaonic atoms \cite{Hirenzaki:2000da,Baca:2000ic} and leads to deep $K^-$ states in medium and heavy nuclei bound by 30-40~MeV with sizable widths $\sim 100$~MeV.
In contrast with the former results, it should be noted the recent analysis in Refs.~\cite{Cieply:2011fy,Friedman:2012qy}. Starting from a $\bar K N$ pseudopotential inspired in the lowest order $s$-wave amplitude from the meson baryon chiral Lagrangian, constrained by data within the energy region $\simeq$30-50~MeV below threshold (as probed by kaonic atom phenomenology), and solving the coupled-channel problem  with in-medium hadron selfenergies, a deeper potential about twice the size of previous determinations is obtained (i.e. $V_{K^-}\simeq -80$~MeV at threshold). A subsequent fit to kaonic atom data including additional quadratic terms in the potential (mimicking absorptive channels not included in the single-nucleon interaction) increases the discrepancy by a factor of three. There are, however, important differences regarding the zero-energy limit of the proposed scattering potential, the double-pole structure of the emerging $\Lambda(1405)$ state, the treatment of the (off-shell) energy dependence of meson selfenergies and the self-consistent implementation of the kaon potential, as compared to previous determinations (see, e.g., Refs.~\cite{Ramos:1999ku,Lutz:2007bh}), which may explain the differences found at the level of the one-body interaction and the stronger contribution of the two-body mechanisms at central nuclear densities.

The properties of both $K$ and $\bar K$ close to threshold have also been thoroughly investigated in HICs at SIS energies \cite{Fuchs:2005zg,Forster:2007qk,Hartnack:2011cn,Cassing:2003vz}. The analysis of experimental data (such as production cross sections, energy and polar angle distributions) in conjunction with microscopic transport approaches has allowed to draw solid conclusions regarding the production mechanisms of strangeness, the different freeze-out conditions exhibited by $K^+$ and $K^-$ mesons and the use of $K^+$ as a probe of the nuclear matter equation of state at high densities. A good agreement with data has been achieved for many different observables when the in-medium properties of kaons are implemented. Still, a consensus on the influence of the kaon-nucleus potential on the whole set of experimental data involving antikaon production is still missing, leaving room for a more elaborate description of the most relevant reactions (e.g., $\pi Y \to \bar K N$) within hadronic models.

An early motivation for the present study of the properties of strange vector mesons originates in the substantial experimental activity within the RHIC low energy scan programme and the HADES experiment at GSI, which are currently performing measurements in order to extract the in-medium properties of hadronic resonances and, in particular, of strange vector mesons from HICs \cite{Blume:2011sb,Agakishiev:2013nta}. Recent developments within the last 10-15 years suggest the use of hadronic many-body models together with microscopic transport approaches (particularly those incorporating off-shell effects) as a powerful tool to investigate the complicated reactions that take place in HICs.
Within this scenario strange vector meson resonances have triggered attention only recently, mostly due to the fact that, unlike their unflavoured partners, they do not decay in the dilepton channel, making their experimental detection less clean. On the theory side, the properties of $\bar K^*$ have been studied in dense nuclear matter \cite{Tolos:2010fq} starting from a model of the $\bar K^* N$ interaction within the hidden gauge formalism. In a similar way as for the $\bar K$ meson, the excitation of several $YN^{-1}$ components leads to a large broadening of the $\bar K^*$ spectral function with a moderate attraction on the quasiparticle energy. The $K^*$, however, has not been studied within this framework. Presumably, the lack of baryonic resonances with positive strangeness calls for much more moderate medium effects in the $S=+1$ sector. Still, such information is
important for the present experimental program at GSI and for the
future FAIR facility. The present study contains novel
aspects which will cover part of the missing information at the theoretical level.

In the context of hot hadronic matter with low baryonic content some estimations have been done for the $\phi$ meson decay at temperatures close to deconfinement based on the chiral dynamics of kaons and anti-kaons in a hot gas of pions \cite{Faessler:2002qb}. The same scenario with the strange vector mesons $K^*$ and $\bar K^*$ is, however, unexplored.
In heavy-ion collisions at RHIC and
LHC energies the matter at midrapidity is practically baryon free
and dominated in the later stages by mesons. Presently, there is a big interest in the
experimental study of $K^*$, $\bar K^*$ production at ultrarelativistic
energies since the vector strange resonances are considered as a
promising probe of the hot medium and the quark-gluon plasma (QGP) formation - cf.~e.g.~the review [15].
Therefore having a theoretical asset on the properties of strange vector mesons in hot matter is also very important for the interpretation of the experimental
observations. We conclude from our study that 
the $K^*$ and $\bar K^*$ change their
properties only moderately in hot matter, leaving the hope that the
strange vector mesons might be used as a promising probe of QGP
properties since they are not strongly distorted by final-state
hadronic interactions.

With the focus on the implementation of effective hadronic models based on chiral dynamics in microscopic transport simulations \cite{Cassing:2003vz,Cassing:1999wx}, we present in this paper a study of the spectral function of strange vector mesons within a Breit-Wigner quasiparticle picture, where the effects of the nuclear environment are encoded in density or temperature dependent masses and decay widths. An important caveat worth mentioning is the role of pionic medium effects. Whereas these have been extensively studied in dense baryonic matter within hadronic many body theory, the complexity of microscopic transport simulations makes the implementation of pion in-medium properties an enormous task, since pions participate in many different reactions, either as product or in the initial state. In addition, the current precision of experimental data on observables from HICs makes it difficult to resolve whether the role of pion in-medium properties is a determinant factor. In this respect, we constrain ourselves in the present work to the medium effects on strangeness, and we shall assume that pions are on-shell quasiparticles with vacuum-like properties in all explicit calculations involving $\bar K^* / K^*$ decays.

This work is organised as follows. In Section~\ref{sec:cold-matter} we introduce the Breit-Wigner approach for the spectral function and discuss our results for dense nuclear matter in terms of collisional and decay width effects, emphasising the differences between the $K^*$ and the $\bar K^*$. We present a calculation of the $K^*$ selfenergy within the hidden local symmetry formalism whereas for the $\bar K^*$ we build upon previous results in the same approach. The high-temperature scenario is explored in Sec.~\ref{sec:hot-matter} by implementing kaon dynamics in a pion gas from a chirally constrained model in the decays of $K^*/\bar K^*$s. We also study whether an in-medium mass shift of the $K^*$ at finite temperature appears from the dispersion relation of the in-medium energy dependent $K^*$ decay width. Finally, we draw our summary and conclusions in Sec.~\ref{sec:summary}.

  \section{Medium effects in dense nuclear matter}
\label{sec:cold-matter}

In the present study we adopt the relativistic Breit-Wigner prescription for the strange meson in-medium spectral function
\cite{Bratkovskaya:2007jk},
  \begin{align}
    A_{i} (M,\rho_{N}) = C_{1} \, \frac{2}{\pi} \frac{M^{2} \Gamma_{i}^{*} (M,\rho_{N})}{\left(M^{2} - {M_i^{*}}^{2} (\rho_{N}) \right)^{2} + \left( M \Gamma_{i}^{*} (M,\rho_{N}) \right)^{2}} \ , 
  \label{eq:bwsf} 
  \end{align}
where $C_1$ stands for a normalisation constant, which is determined as the spectral function must fulfil the sum rule $\int_0^{\infty}A_{i} (M, \rho_{N}) \SPC dM = 1$, and $i=K/\bar K, K^*/\bar K^*$ throughout this work. In Eq.~(\ref{eq:bwsf}) we have explicitly written the dependence on the nuclear matter density $\rho_N$ to indicate medium effects. In Sec.~\ref{sec:hot-matter} this will be replaced by a dependence on the temperature of the thermal bath.
We also note that in general the strange meson width $\Gamma_i^*$ depends on the (off-shell) energy $M$, thus Eq.~(\ref{eq:bwsf}) departs from the standard Lorentzian parameterisation used in quantum mechanics but keeps closer to the actual phenomenology of strange vector meson resonances. In the quantum-field theory sense, the latter is mandatory since the decay width of a bosonic state is tied to the imaginary part of the associated selfenergy, namely $\Gamma_i^*(M,\rho_N) = -\textrm{Im} \, \Pi_i(M,\rho_N) /M$. The real part of the selfenergy is often reabsorbed in the definition of the in-medium meson mass, denoted by $M_i^*$ above, which is a good approximation as long as $\delta M_i = M_i^*-M_i \ll M_i$ (with $M_i$ being the nominal mass in vacuum) and $\textrm{Re} \, \Pi_i$ is not strongly energy dependent. To be precise, the in-medium meson mass is defined as the solution of the meson dispersion relation at vanishing momentum,
  \begin{align}
  (M_i^*)^2-M_i^2-\textrm{Re}\,\Pi_i(M_i^*,\rho_N)=0 \ .
\label{eq:mass-shift}
  \end{align}
The approximations discussed above typically introduce small violations of the spectral function sum rule which is why we explicitly keep a normalisation constant.

It is also worth mentioning that, in general, the meson selfenergy operator depends \emph{separately} on the energy and momentum of the particle, as a consequence of the breaking of Lorentz invariance induced by the existence of a preferred reference frame, that of the nuclear environment. The parameterisation in Eq.~(\ref{eq:bwsf}) does not take this fact into account. As we shall discuss below, we explore different implementations of the in-medium properties based on previous model calculations of the selfenergy of strange mesons. As for the momentum dependence, this limitation in our parameterisation will be overcome by considering static properties, i.e. by assuming the strange meson at rest with respect to the nuclear medium.
Similarly, the vector nature of $\bar K^* / K^*$ mesons induces a splitting of their spectral function in longitudinal ($L$) and transverse ($T$) modes. We shall restrict ourselves to the case of zero momentum where both the $L$ and $T$ spectral functions become degenerate.

We distinguish two sources of medium effects that contribute to the selfenergy of a strange meson in a nuclear medium: (i) the modification of its dominant decay mode, e.g.~$\bar K^* \to \bar K \pi$, induced by medium effects on the light pseudoscalars (also typically referred to as two-meson cloud effects); and (ii) the quasi-elastic interaction of the strange meson with nucleons, e.g.~$\bar K^{(*)} N \to \bar K^{(*)} N$ and related absorptive channels. Obviously the first mechanism is only present for strange vector mesons, which have predominant decay modes into $\bar K (K) \pi$. In the same fashion we write the in-medium width $\Gamma^*$ entering Eq.~(\ref{eq:bwsf}) as $\Gamma^*=\Gamma_{\textrm{dec}}+\Gamma_{\text{coll}}$, the sum of the width associated to the main decay mode ($\Gamma_{\textrm{dec}}$) plus a \emph{collisional} width ($\Gamma_{\text{coll}}$) emerging from the scattering with nucleons. The latter will be either parameterised from existing models or else calculated explicitly, whereas the former one can be evaluated straightforwardly as \cite{Bratkovskaya:2008iq}
 \begin{align}
    \Gamma_{V,\, \textrm{dec}} (\mu, \rho_{N}) = \Gamma_{V}^{0} \left( \frac{\mu_{0}}{\mu} \right)^{2} \frac{\int_{0}^{\mu - m_{\pi}} q^3 (\mu, M) A_{j}(M,\rho_{N}) \SPC dM}{\int_{M_{\rm min}}^{\mu_{0} - m_{\pi}} q^3 (\mu_{0}, M) A_{j} (M,0) \SPC dM \ },\label{eq:vmdw} 
  \end{align}
where $j = K, \bar{K}$, $q(\mu, M) = \sqrt{\lambda(\mu, M, M_{\pi})}/ 2 \mu$, $\lambda(x,y,z) = \left[ x^{2} - (y + z)^{2} \right] \left[ x^{2} - (y - z)^{2} \right]$ and we have explicitly included the subindex $V$ as this expression stands for the ($p$-wave) decay of the strange vectors $\bar K^*$ and $K^*$. 
$\Gamma_{V}^{0}$ stands for the vector meson (vacuum) partial decay width in the considered channel and $\mu_0$ is the nominal resonance mass, particularly $\Gamma_{K^{*},\bar{K}^{*}}^{0} = 42$~MeV and $\mu_0=892$~MeV \cite{Beringer:1900zz}.
We note that the same expression can be derived from the imaginary part of the lowest order selfenergy diagram in Fig.~\ref{fig:ksloop} \cite{Tolos:2010fq,Cabrera-proc}.
Eq.~(\ref{eq:vmdw}) accounts for the in-medium modification of the resonance width by its decay products. In particular, we consider the fact that kaons and anti-kaons may acquire a broad spectral function in the medium, $A_{j}(M,\rho_N)$. Pions will be assumed to stay as narrow quasiparticles with vacuum properties in the evaluation of $\Gamma_{V,\, \textrm{dec}}$ throughout this work.
This choice is, most likely, not realistic at finite nuclear density (we refer, for instance, to the classic references \cite{Ericson:1988gk,Oset:1981ih}), whereas in hot matter with low baryonic content the pion is expected to experience small changes up to temperatures $T\simeq 100$~MeV \cite{Goity:1989gs,Schenk:1993ru}. We want to focus on the role of medium effects on the strangeness degrees of freedom, while sticking to the typical set-up of transport calculations. Thus our results for $\Gamma_{V,\,{\rm dec}}$ should be considered as a lower boundary and reference to other calculations accounting for pion properties will be made where possible.
Finally we note that, for the case of interest, $A_{j}(M,0)$ in Eq.~(\ref{eq:vmdw}) is a delta function since the kaon is stable in vacuum with respect to the strong interaction. In general this may not be the case, e.g., $a_1\to \rho\pi$, where the $\rho$ meson decays into two pions most of the time ($M_{\rm min}$ then stands for the threshold energy).

\subsection{$K^{*}$ meson}

The absence of baryonic resonances with $S=+1$ close to threshold induces milder nuclear medium effects in the properties of the $K$ meson \cite{Ramos:1999ku,Tolos:2008di} as compared to the $\bar K$ meson, whose behaviour is largely dominated by the $\Lambda (1405)$ resonance appearing in $s$-wave $\bar K N$ scattering.
A similar situation is expected to take place for the vector partner of the $K$, the $K^*$ meson. Note that the $K^*$ decays predominantly into $K \pi$, and therefore not only collisional effects but the in-medium decay width has to be taken into account. However, since the $K$ meson itself is barely influenced by nuclear matter one can anticipate small changes in $\Gamma_{K^*,\,{\rm dec}}$ as compared to the vacuum case.

\begin{figure}
  \centering
    \includegraphics[width=.7\linewidth]{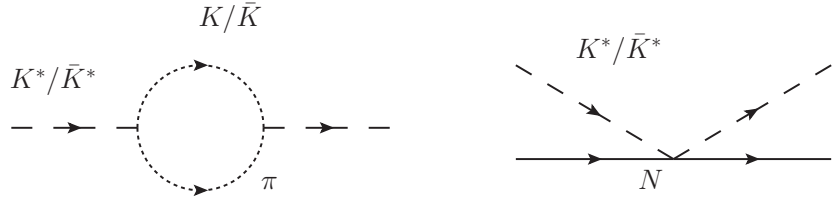}
    \caption{Left: The $K^* (\bar K^*) \to K (\bar K) \pi$ loop selfenergy diagram. Right: Tree level diagram for elastic scattering between a $K^* (\bar K^{*})$ and a nucleon.}
    \label{fig:ksloop}
  \end{figure}

The collisional selfenergy of the $K^*$ has not been evaluated before in the context of effective theories of QCD. We take recourse of the hidden local symmetry approach in \cite{Oset:2009vf}, which has been recently used to study the selfenergy of the $\bar K^*$ in cold nuclear matter \cite{Tolos:2010fq} (see \cite{Oset:2012ap} for a recent review). In this approach, the interaction of light vector mesons with the $J^P=1/2^+$ baryon octet is built by assuming that the low-energy amplitudes are dominated by vector meson exchanges between the baryons and the vector themselves, which couple to each other. At small momentum transfer, the transition potential (tree-level scattering amplitude, c.f.~Fig.~\ref{fig:ksloop}) is completely determined by chiral symmetry breaking. Neglecting corrections of order $p/M$, with $p$~$(M)$ the baryon three-momentum (mass), the $s$-wave projection reads
\begin{eqnarray}
\label{eq:VB-potential}
V_{ij}
&=& - C_{ij} \frac{1}{4f^2} (2\sqrt{s}-M_{B_i}-M_{B_j}) \,
\left ( \frac{M_{B_i}+E_i}{2 M_{B_i}} \right )^{1/2} \left ( \frac{M_{B_j}+E_j}{2 M_{B_j}} \right )^{1/2}  \,\vec{\varepsilon}\cdot\vec{\varepsilon}\,' \nonumber \\
&\simeq& - C_{ij} \frac{1}{4f^2}(q^0+q'^0) \,\vec{\varepsilon}\cdot\vec{\varepsilon}\,' \ ,
\end{eqnarray}
where $q^0$~($q'^0$) stands for the energy of the incoming~(outgoing) vector meson with polarisation $\vec{\varepsilon}$~($\vec{\varepsilon}\,'$), $C_{ij}$ stand for channel-dependent symmetry coefficients \cite{Oset:2009vf}, and the latin indices label a specific vector-meson baryon ($VB$) channel, e.g., $K^{*+} p$. For practical purposes the second equation is satisfied to a good approximation.
Within this approach the $K^*$ selfenergy can be evaluated straightforwardly in a $t\rho$ approximation (justified for the present study)\footnote{The normalisation of the scattering amplitude $T$ (or $V$) is such that the differential cross section reads $\frac{d\sigma}{d\Omega} = \frac{1}{16\pi^2}\frac{MM'}{s}\frac{q'}{q} \,|T|^2$.},
  \begin{align}
    \Pi^{\rm coll}_{K^{*}} = \frac{1}{2} \left( V_{K^{*+} p} + V_{K^{*+} n} \right) \rho_{0}
    \left(\frac{\rho_{N}}{\rho_{0}}\right) 
    =
    \frac{1}{4} (V_0+3V_1) \rho_{0}
    \left(\frac{\rho_{N}}{\rho_{0}}\right) \ ,
  \label{eq:Kstar-trho}
  \end{align}
where in the second equation $V_I$ stands for the potential in isospin basis and we omit the polarisation vectors for simplicity of notation.
By taking $q^0=q'^0\simeq M_{K^*}$, and using $C_{K^{*+} p} = - 2$, $C_{K^{*+} n} = - 1$ \cite{Oset:2009vf} and $f=93$~MeV we obtain
\begin{align}
\Pi^{\rm coll}_{K^{*}} \simeq \alpha \frac{M_K}{M_{K^*}} M_{K^*}^2 \left(\frac{\rho_{N}}{\rho_{0}}\right) \ ,
\label{eq:Kstar-trho-2}
\end{align}
with $\alpha=0.22$, leading to a positive mass shift (equivalent to a repulsive optical potential) of about $\delta M_{K^*}\simeq 50$~MeV at normal matter density ($\rho_0=0.17$~fm$^{-3}$).

Unitarisation of the leading order meson baryon amplitudes based on the chiral Lagrangian has been used in the strange sector resulting in a more realistic description of low energy observables such as scattering lengths and cross sections. Particularly, for $S=-1$ it also leads to the dynamical generation of the $\Lambda (1405)$ in $\bar K N$ scattering, thus extending the applicability of the low-energy theory. A similar situation was encountered in $\bar K^* N$ scattering \cite{Tolos:2010fq}, as we shall discuss in the next section. In the $S=+1$ sector, the use of unitarised amplitudes also results in a more reliable description of $KN$ scattering data and optical potential \cite{Tolos:2008di,Oset:1997it,Nekipelov:2002sd}. We follow the same reasoning here and together with the Born approximation, Eq.~(\ref{eq:Kstar-trho}), we present results for the $K^*$ selfenergy by iterating the leading order potential in a
Bethe-Salpeter equation to produce the full scattering amplitude, schematically $T = V+VGT$.
We note that the solution of the Bethe-Salpeter equation is particularly simplified within the chiral effective theory since both the potential $V$ and the resummed amplitude $T$ can be factorised on-shell, and thus the solution proceeds by algebraic inversion. It has been shown that the off-shell parts within the integral term of the equation lead to singularities that can be renormalised by higher-order counterterms and are effectively accounted for by using physical masses and coupling constants. We refer to \cite{Oller:1997ng,Oller:2000fj} for details about different unitarisation schemes in meson-meson and meson-baryon scattering within Chiral Perturbation Theory.

The $K^{*+}p$ state is pure isospin $I=1$ and gets single channel unitarisation, whereas the $K^{*+}n$ amplitude is coupled with the $K^{*0}p$ state and thus a 2$\times$2 matrix problem arises. After some basic algebra, the unitarised amplitudes read
\begin{eqnarray}
\label{eq:KstarN-uni}
T_{K^{*+}p} =  \frac{V}{1-VG} \ , \quad
T_{K^{*+}n} = T_{K^{*0}p} = T_{K^{*0}p,K^{*+}n} = \frac{1}{2} \frac{V}{1-VG} \ , 
\end{eqnarray}
with $V=(q^0+q'^0)/2f^2=V_{K^{*+} p}$. $G$ stands for the meson-baryon loop function, which can be found, for instance, in \cite{Oset:1997it}. We adopt here the same regularisation scale for $G$ as it was fixed in $\bar K N$ scattering studies: in a cut-off scheme this corresponds to $q_{\rm max}\simeq630$~MeV$/c$ \cite{Oset:1997it} or, equivalently, to a subtraction constant of natural size $a_{K^* N}\simeq -2$ in dimensional regularisation \cite{Oller:2000fj,Jido:2002zk}.
Replacing the lowest order tree level amplitudes in Eq.~(\ref{eq:Kstar-trho}) by the former expressions one finds a reduction by roughly one third over the previous result, namely $\delta M_{K^*}(\rho_0)\simeq 30$~MeV [i.e. $\alpha\simeq 0.13$ in Eq.~(\ref{eq:Kstar-trho-2})].

\begin{figure}
    \centering
    \includegraphics[width=0.6\textwidth]{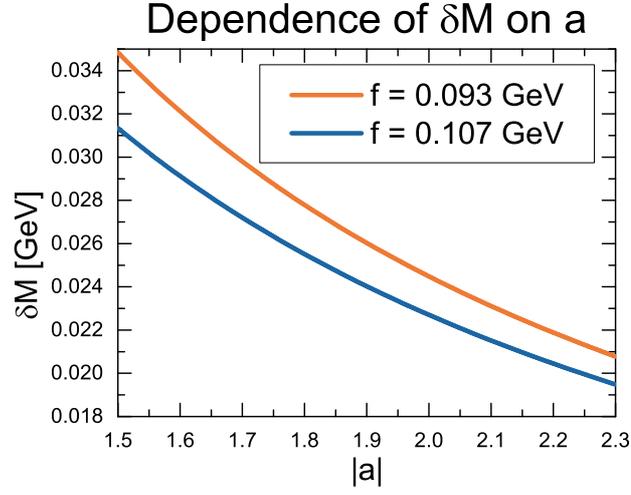}
    \caption{$K^{*}$ mass shift from Eqs.~(\ref{eq:mass-shift}, \ref{eq:Kstar-trho-2}, \ref{eq:KstarN-uni}) at normal matter density as a function of the subtraction constant in the meson-baryon loop function, for two values of the pseudoscalar decay constant $f$.}
    \label{fig:f-and-a-dep}
  \end{figure}

Several comments regarding the chosen values for $f$ and $a_{K^* N}$ are in order.
First, previous experience from $\bar K (K) N$ studies indicates that the agreement with meson-baryon scattering data improves when leaving $f$ as a free parameter, and typically an intermediate value between $f_{\pi}$ and $f_{K}$ as obtained in meson-meson calculations comes out from these analysis \cite{Oset:1997it}, $f_K \simeq 1.15 f_{\pi}=107$~MeV. Note, however, that this responds to a strictly phenomenological motivation in the pseudoscalar sector, where the amount of data available permits to constrain further the model\footnote{We note, for instance, the study of \cite{Oller:2000fj} where a rather good agreement with $\bar K N$ cross sections and threshold ratios were obtained by using $f=f_0=86.4$~MeV (corresponding to the LO perturbative result in $SU(3)$ ChPT) and a slightly different subtraction constant.}.
In the sector of strange vector meson interactions with baryons, the amount of experimental data is comparatively much smaller, and insufficient to simultaneously constrain both $f$ and the subtraction constant $a_{K^* N}$ (which in fact are substantially correlated). In absence of further knowledge one sticks to $f=93$~MeV, which in the Hidden Local Symmetry formalism leads to vector self-interactions respecting the KSFR rule and vector meson dominance \cite{Oset:2009vf}. Still, one may expect a similar  renormalisation of $f$ here and we shall estimate this uncertainty by calculating the $K^*$ selfenergy with both $f=93,107$~MeV.
Second, our choice of $a_{K^* N}\simeq -2$ responds to former determinations in the $\bar K N$ system which also provide a fair description of scattering observables in the $S=+1$ sector such as $I=0,1$ scattering lengths and low-energy phase shifts\footnote{See also the recent analysis in \cite{Ikeda:2012au} where other possible values of $a_{MB}$ in coupled channels accommodate the data in the $S=-1$ sector. Incidentally the reported value of $a_{\bar K N}$ is not far from $\simeq -2$, and no statement is done for the $S=+1$ sector.}. Lacking further constraints on the subtraction constant in the vector meson sector, it is also pertinent to study the uncertainty in the $K^*$ selfenergy due to this parameter. In order to have an educated estimate we take recourse of previous results in $\bar K (K) N$ calculations.
The value of the subtraction constant cannot be arbitrarily changed as it is correlated to the momentum scale running in the meson baryon loop (with a characteristic exponential dependence \cite{Oller:2000fj}), so that smaller values of, e.g., $a_{KN}$ in absolute magnitude correspond to higher values of $\mu\equiv q_{\rm max}$ (in the limiting case $a_{KN}=-2\log2\simeq -1.39$ one has $\mu\to\infty$). A ``natural'' value window around $-2$ is provided by setting $\mu$ in the region of the first (vector) meson resonance, the $\rho(770)$, whereas the ``infrared'' boundary is better constrained from phenomenology.
By using our unitarised $s$-wave scattering amplitude for the $KN$ system we have obtained a limiting range of variation for $a_{KN}$ of $[-2.3,-1.5]$, by comparing to the total (elastic) $K^+p$ cross section (below $q_{\rm LAB}\simeq 800$~MeV$/c$ the reaction is elastic and therefore coupled-channel dynamics does not play a role in this case).
Comparison to the $I=1$ $KN$ phase shifts close to threshold provides more stringent constraints and best results are obtained within the interval $a_{KN}=-1.8\pm0.1$ (corresponding to $700$~MeV$/c \lesssim q_{\rm max} \lesssim 1000$~MeV$/c$), slightly above $-2$ as anticipated in \cite{Kaiser:1995eg,Oset:1997it}. Particularly, we obtain for the $I=1$ scattering length ${\cal A}_{KN}^{I=1} = -0.31$~fm, in good agreement with experiment and previous determinations.
In Fig.~\ref{fig:f-and-a-dep} we summarise our estimation of uncertainty in the $K^*$ selfenergy by plotting the $K^*$ mass shift at normal nuclear matter density as a function of the subtraction constant for two values of $f$. Within errors we estimate a small, positive mass shift for the $K^*$ around 5\% at $\rho_N=\rho_0$.

Finally, the decay width of the $K^*$ is evaluated according to Eq.~(\ref{eq:vmdw}), where we adopt a narrow quasiparticle approximation for the $K$ spectral function, namely
\begin{equation}
\label{eq:Kspec}
A(M,\rho_N) = 2M\,\delta \lbrack M^2-M_K^{*\, 2}(\rho_N) \rbrack
\end{equation}
with $M_K^{*\, 2}(\rho_N) = M_K^2 + \Pi_K(\rho_N)$ and the $K$ selfenergy as determined in \cite{Oset:1997it,Kaiser:1995eg} within a $t\rho$ approximation, $\Pi_K(\rho_N) \simeq 0.13 M_K^2 (\rho_N/\rho_0)$, and consistent with the findings of \cite{Tolos:2008di}. As expected, the $K^*$ in-medium width is slightly reduced at finite density as the phase space for the $KN$ system is quenched due to the larger $K$ mass.
 
\begin{figure}
    \centering
    \includegraphics[width=0.6\textwidth]{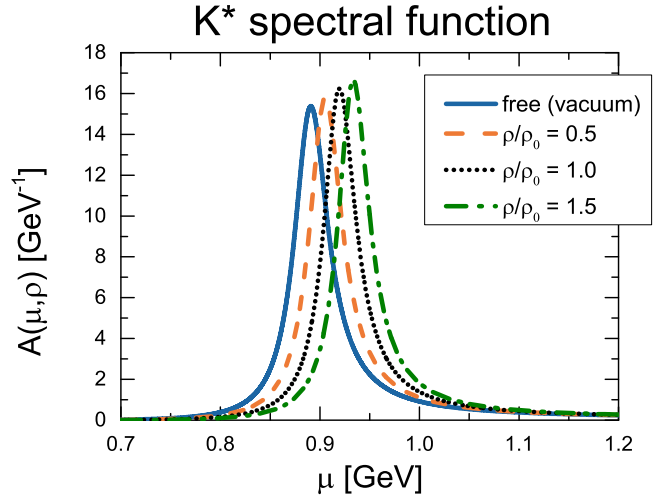}
    \caption{The Breit-Wigner spectral function of the $K^{*}$ as a function of energy for different baryonic densities, including the collisional mass shift from $K^* N$ scattering ($f=107$~MeV, $a_{K^*N}=-1.8$) and $K \pi$ width with in-medium kaons.}
    \label{fig:dkspsf}
  \end{figure}

The spectral function of the $K^*$ is depicted in Fig.~\ref{fig:dkspsf}, according to Eq.~(\ref{eq:bwsf}) with the medium effects discussed above. At normal matter density the $K^*$ quasiparticle peak experiences a repulsive shift corresponding to the mass modification in Eqs.~(\ref{eq:Kstar-trho},\ref{eq:Kstar-trho-2},\ref{eq:KstarN-uni}). In spite of a nominal reduction in $\Gamma_{K^*,\,{\rm dec}}$ due to the increase of $M_K$ at finite density, the shift in the $K^*$ mass increases the available energy for the $K \pi$ decay and the $K^*$ spectral function exhibits practically negligible changes in shape.
We recall that we have explicitly neglected the effect of pion modifications in the $K^*$ width, which may lead to changes up to a factor of $\simeq$2 at normal matter density \cite{Cabrera-proc}.
Overall the $K^*$ behaves as a quasiparticle with a single-peak distribution function and a modified effective mass as a result of the repulsive $K^* N$ interaction.
It is worth mentioning that the HADES Collaboration have recently reported a reduction of the $K^{*0}$ yield in Ar$+$KCL collisions at 1.76 $A$GeV as compared to estimations within the UrQMD transport approach \cite{Agakishiev:2013nta}.

\subsection{$\bar{K}^{*}$ meson}

The properties of $\bar K^*$ have been studied in cold nuclear matter \cite{Tolos:2010fq} starting from a model of the $\bar K^* N$ interaction within the hidden gauge formalism, including coupled channels and imposing unitarisation of the scattering amplitudes, as introduced in the previous section. The set of coupled equations determining the $\bar K^* N$ scattering amplitude and the $\bar K^*$ selfenergy (schematically $\Pi_{\bar K^*}=\sum_{\vec{p}} n(\vec{p}\,)\, T_{\bar K^* N}$) were solved selfconsistently accounting for Pauli-blocking effects, baryonic mean-field potentials and the selfenergy of intermediate meson states. Such scheme is usually known as $G$-matrix approach in reference to the in-medium effective $T$-matrix obtained in Dirac-Brueckner theory.

The collisional part of the selfenergy, related to the quasi-elastic reaction $\bar K^* N \to \bar K^* N$ and accounting for absorption channels, induces a strong broadening of the $\bar K^*$ spectral function as a result of the mixing with two $J^P=1/2^-$ states, the $\Lambda(1783)$ and $\Sigma(1830)$, which are dynamically generated in a parallel way to the $\bar K N$ interaction. At $\rho_N=\rho_0$, it was found that the $\bar K^*$ width is enlarged beyond 200~MeV, the in-medium $\bar K \pi$ decay mode contributing about 100~MeV. The $\bar K^*$ nuclear optical potential indicates a moderate attraction of about 50~MeV. The quasiparticle peak, however, exhibits a larger attraction as a consequence of the interference with the collective modes above threshold, which are responsible for a considerable energy dependence of the $\bar K^*$ selfenergy. This reflects the limitations of the quasiparticle representation of the spectral function of the system, which we illustrate further in the following.

To construct the spectral function of the $\bar{K}^{*}$ according to Eq.~(\ref{eq:bwsf}) we have obtained the effective in-medium mass and width by solving the $\bar K^*$ dispersion relation utilising the results for the $\bar K^*$ selfenergy obtained in \cite{Tolos:2010fq}, i.e., $\omega_{\bar K^*}^2-M_{\bar K^*}^2-\textrm{Re}\,\Pi_{\bar K^*}(\omega_{\bar K^*},\rho_N)=0$ and $\Gamma^*_{\bar K^*}(\rho_N)=-\textrm{Im} \, \Pi_{\bar K^*}(\omega_{\bar K^*},\rho_N) /M_{\bar K^*}$ (no energy dependence is retained in $\Gamma^*_{\bar K^*}$). We shall denote this method as ``G-matrix~1''. We note that the results of \cite{Tolos:2010fq} also account for the in-medium decay width for $\bar K^*\to \bar K \pi$, thus our parameterisation according to the dispersion relation with the total selfenergy captures both the collisional and the decay width of the $\bar K^*$ at the quasiparticle energy.
We recall that the $\bar K^*$ meson spectral function in a nuclear medium depends explicitly on energy and momentum \cite{Tolos:2010fq},
  \begin{align}
 S_{\bar K^*} (\omega, \vec{q}; \rho_{N}) = - \frac{1}{\pi} \frac{\textrm{Im}\,\Pi_{\bar K^*} (\omega,\vec{q};\rho_{N})}{\left[ \omega^{2} - \vec{q}\,^{2} -  M_{\bar K^*}^{2} - \textrm{Re}\,  \Pi_{\bar K^*} (\omega, \vec{q}; \rho_{N})  \right]^{2} + \left[ \textrm{Im}\,  \Pi_{\bar K^*} (\omega, \vec{q}; \rho_{N}) \right]^{2}}.
  \end{align}
The proposed parameterisation, relying on Eq.~(\ref{eq:bwsf}), is only suitable for $\vec{q}=0$, namely
  \begin{align}
    A_{\bar K^*} (M,\rho_{N}) = 2 \, C_{1} \, M \, S_{\bar K^*} (M,\vec{0};\rho_{N}) \ .
  \end{align}

  \begin{figure}
    \centering
    \includegraphics[width=0.5\textwidth]{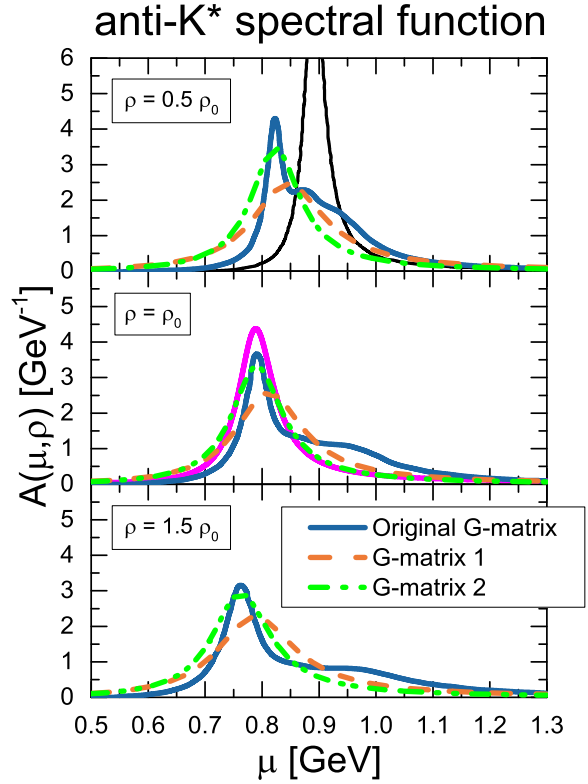}
    \caption{The spectral function of the $\bar{K}^{*}$ for different baryonic densities.  The blue solid line corresponds to the original calculation in \cite{Tolos:2010fq} by Tolos et al. The Breit-Wigner parameterisations with in-medium mass and decay width are shown in orange dashed (``G-matrix 1'') and green dash-dotted (``G-matrix-2'') lines, c.f.~the text for details. The pink solid line in the middle panel corresponds to the explicit evaluation of the $\bar K\pi$ width with in-medium kaons. The vacuum spectral function is also displayed in the upper panel for reference (black thin solid line). The normalisation of the Breit-Wigner functions is set as to match the results of \cite{Tolos:2010fq}.}
    \label{fig:dksmsf}
  \end{figure}

Our results can be seen in Fig.~\ref{fig:dksmsf}, where we compare our parameterisation with the original $\bar K^*$ spectral function from \cite{Tolos:2010fq}. It is evident that a Breit-Wigner form with in-medium properties based on the quasiparticle picture cannot retain the full structure of the original spectral function, particularly the modes appearing at the right hand side of the $\bar K^*$ peak, which are related to the $\Lambda h$ and $\Sigma h$ excitations of the system.
The position of the peak in our result lies close to the $\bar K^*$ quasiparticle peak in the full result, as it is determined by the solution of the dispersion relation, which we account for explicitly. The obtained width, however, is larger than reflected by the full calculation. This is related to the fact that the $\bar K^*$ selfenergy is highly energy dependent in the vicinity of the quasiparticle energy, due to the presence of the $\Lambda(1783) N^{-1}$ and the $\bar K \pi$ thresholds, which induce a quick rise of the $\bar K^*$ width. Such energy dependence is not kept in our parameterisation and the resulting distribution exhibits a larger width which, incidentally, partly makes up for the missing strength in the higher energy region. In the low energy part, however, the broadening of our distribution overestimates the strength of the spectral function from the full calculation.

Alternatively, in order to estimate the effect of the energy dependence in the resonance width, we correct the quasiparticle energies from the dispersion relation so as to match the actual position of the $\bar K^*$ peak, i.e. $\omega_{\bar K^*} \to \tilde{\omega}_{\bar K^*} = \omega_{\bar K^*} + \Delta\omega \ | \ {\rm Im}\,\Pi_{\bar K^*}'(\tilde{\omega}_{\bar K^*},\rho_N)=0$ and $\Gamma^*_{\bar K^*}(\rho_N)=-\textrm{Im} \, \Pi_{\bar K^*}(\tilde{\omega}_{\bar K^*},\rho_N) /M_{\bar K^*}$. We denote this parameterisation method as ``G-matrix~2''. The resulting spectral function, c.f.~Fig.~\ref{fig:dksmsf}, now peaks at the same position as the full calculation and exhibits a smaller width in closer agreement with the $\bar K^*$ structure, the reduction in the width tied to lower quasiparticle energies leading to a more quenched in-medium phase space. Consequently, the upper region containing the strange collective modes is rather underestimated, as expected since we cannot reproduce multi-mode structures with a simple Breit-Wigner parameterisation. Finally, as an additional test we show in the middle panel of Fig.~\ref{fig:dksmsf} (solid purple line) our spectral function with a similar parameterisation of the collisional part of the width, $\Gamma_{\bar K^*,\,\text{coll}}$, and explicitly calculating the $\bar K \pi$ decay width $\Gamma_{\bar K^*,\,\textrm{dec}}$ according to Eq.~(\ref{eq:vmdw}). For this we have taken recourse of the results in \cite{Ramos:1999ku} for the $\bar K$ spectral function at normal nuclear density. The resulting distribution is narrower than any of the former proposed methods, which precisely reflects the contribution of in-medium pions in the $\bar K\pi$ cloud from \cite{Tolos:2010fq}.

We conclude that the complex many-body dynamics present in the $\bar K^*$ spectral function in dense nuclear matter, as a result of the interference between several collective excitations of the system, cannot be successfully reproduced in a Breit-Wigner scheme.
The main features of the full dynamics, such as considerable broadening and attractive shift of the quasiparticle mode, of special importance in transport calculations, can be parameterised in terms of a density dependent effective mass and width according to the dispersion relation in the full microscopic model of \cite{Tolos:2010fq}. Still, other details such as the modification of thresholds in the medium, which strongly influences the low-energy behaviour of the spectral function, are also important from the point of view of transport simulations since they modify the onset of strangeness production reactions (e.g.~$NN\to \bar K X$). Thus for a precise description of these features one is bound to recourse to the results from realistic many-body evaluations.

  \section{Medium effects in hot hadronic matter}
\label{sec:hot-matter}

In this section we focus on the scenario of a hot hadronic system with low net content of baryons, i.~e., $\mu_B \simeq 0$. Then the medium is mostly constituted by thermal excitations of the lightest hadrons, namely pions, and at next to leading order heavier mesons ($K$, $\bar K$, $\eta$,...). Such a set-up corresponds to the late evolution of a high-energy heavy-ion collision, as produced in RHIC and LHC. The $\mu_B\ne 0$ scenario, where the interaction with nucleons sets in at a finite temperature, is expected to be encountered at low and intermediate energy collisions, to be produced in the near future at the FAIR and NICA facilities.

\begin{figure}
    \centering
    \includegraphics[width=0.6\textwidth]{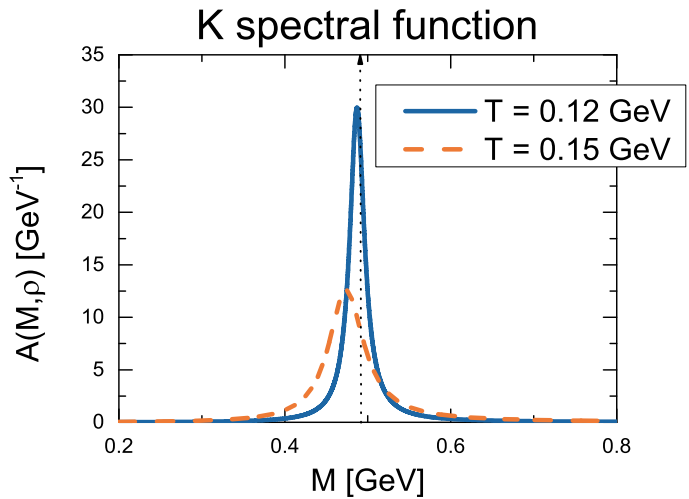}
    \caption{The spectral function of the $K$ is shown as a function of the invariant mass $M$ for different temperatures.  The blue solid line corresponds to $T = 0.12$~GeV and the orange dashed line corresponds to $T = 0.15$~GeV. The vertical dotted line indicates the position of the vacuum kaon mass.}
    \label{fig:tksf}
  \end{figure}

In a hot, isotopically symmetric pionic medium, both the $\bar K^*$ and $K^*$ vector mesons behave identically. The medium effects in this case are tied to the modifications of the predominant decay mode, $\bar K (K) \pi$.
The in-medium properties of kaons in hot matter have been studied by Fuchs et al. \cite{Faessler:2002qb} in the framework of Chiral Perturbation Theory, together with a phenomenological extension to reach high temperatures (up to $T\lesssim M_{\pi}$) based on experimental $K \pi$ phase shifts.
The kaon selfenergy is obtained as the sum of the forward $K \pi$ scattering amplitudes over the pion momentum distribution function, namely
\begin{equation}
\label{eq:Pi-finiteT}
\Pi_K(M_K,T) = - \int T^+_{K\pi}(s,0,u) (dn_{s\pi^+}+dn_{s\pi^0}+dn_{s\pi^-}) + \int T^-_{K\pi}(s,0,u) (dn_{s\pi^+} - dn_{s\pi^-}) \ ,
\end{equation}
where
$$dn_{s\pi} = \frac{d^3p}{(2\pi)^3} \, \frac{1}{2 \,\omega_{\pi}(p)} \, \frac{1}{\exp \left( \frac{\omega_{\pi}(p)-\mu_{\pi}}{T}\right) - 1}$$
is the scalar pion density and $T^{\pm}_{K\pi}(s,t,u)$ are related to the amplitudes in isospin basis $T^{I}_{K\pi}$ as $T^{1/2}_{K\pi} = T^{+}_{K\pi}+ 2 \,T^{-}_{K\pi}$, $T^{3/2}_{K\pi} = T^{+}_{K\pi} - T^{-}_{K\pi}$.
It was found in \cite{Faessler:2002qb} that the kaon experiences a moderate broadening and a comparatively small, attractive mass shift up to $T=100$-$120$~MeV, reaching values of about $\Gamma^*_K \simeq 50$~MeV and $\delta  M_K \simeq -35$~MeV at $T=150$~MeV, which is the highest temperature that we considered here.
By parameterising the results from \cite{Faessler:2002qb} and using Eq.~(\ref{eq:bwsf}) we obtain a kaon spectral function with a single quasiparticle structure that is mildly shifted to lower energies while becoming substantially  broadened only at very high temperatures, c.f.~Fig.~\ref{fig:tksf}.

Next we make use of Eq.~(\ref{eq:vmdw}) to calculate the $K^{*}$ (or $\bar K^*$) decay width, related to the imaginary part of the selfenergy. We note that in this case the collisional broadening is already automatically accounted for since all the relevant dynamics is tied to the coupling to $\bar K (K) \pi$.
Since there is no further information available on a possible mass shift of the $K^*$ in the hot medium, we evaluate it explicitly from the real part of the selfenergy, which we calculate using a dispersion relation over the imaginary part, namely:
  \begin{align}
  \label{eq:DR}
    \textrm{Re}\, \Pi_{K^*}(\mu,T)  = \frac{2}{\pi} \int_{M_{\pi}}^{\infty} d\mu' \, 
    \frac{\mu'}{\mu'^{2} - \mu^{2}} \, \textrm{Im}\,  \Pi_{K^*} ( \mu',T ) \ ,
  \end{align}
where we have renamed the vector meson energy as $\mu$ to avoid confusion with the kaon energy, $M$.
The integral in Eq.~(\ref{eq:DR}) is quadratically divergent and needs to be regularised. This has been done, for instance, in \cite{Klingl:1996by} for neutral vector mesons by using a twice subtracted dispersion relation. Since in the strange sector we lack enough constraints on the subtraction constants we choose instead to remove the vacuum decay width of the $K^*$,
  \begin{align}
  \label{eq:DR-reg}
    \textrm{Re}\, \Pi_{K^*} ( \mu,T )  -  \textrm{Re}\, \Pi_{K^*} ( \mu,0 ) = - \frac{2}{\pi} \int_{M_{\pi}}^{\infty} d\mu' \, \frac{\mu'^{2}}{\mu'^{2} - \mu^{2}} \left[ \Gamma^*_{K^*,\, dec} ( \mu', T ) - \Gamma^{\rm vac}_{K^*} ( \mu' ) \right] \ ,
  \end{align}
which then renders the net medium effect on the physical mass of the particle (the real part of the vacuum selfenergy renormalises the \emph{bare} mass and is not related to medium effects). This is, however, still not enough to guarantee convergence and thus we introduce a phenomenological form factor of dipole form, associated to the $K^* K \pi$ vertex,
  \begin{align}
  \label{eq:ffactor}
    F(\Lambda, \mu,M) = \left( \frac{\Lambda^{2} + q^2(M_{K^{*}}, M_{K})}{\Lambda^{2} + q^{2}(\mu, M)} \right)^2
  \end{align}
which enters the calculation of the $K^*$ decay width in vacuum and in the medium,
  \begin{eqnarray}
  \label{eq:Kstar-widths}
    \Gamma^*_{K^*} \equiv \Gamma_{K^*,\, {\rm dec}} \left( \mu, T \right) &=& \Gamma^{0} \left( \frac{\mu_{0}}{\mu} \right)^{2} \frac{1}{q(\mu_{0}, M_{K})^{3}} \int_{0}^{\mu - m_{\pi}} dM \, A(M,T)\,  q^3(\mu,M) F^2(\Lambda, \mu, M) \ , \nonumber \\
    \Gamma^{\rm vac}_{K^*} \left( \mu \right) &=& \Gamma_{K^*}^{0} \left( \frac{\mu_{0}}{\mu} \right)^{2}  \frac{q^3(\mu, M_{K})}{q^3(\mu_{0}, M_{K})}  F^2(\Lambda, \mu, M_{K}) \ ,
  \end{eqnarray}
and cuts off high-energy dynamics which is not expected to be described by Eq.~(\ref{eq:vmdw}).
The form factor employed here was used in \cite{Bratkovskaya:2007jk,Rapp:1995zy} in the context of $\rho \to\pi\pi$ dynamics in order to improve the comparison with data in vector-isovector $\pi\pi$ scattering (pion electromagnetic form factor). The $\rho$ decay into $\pi\pi$ probes similar kinematics as the $K^*$ which is why we choose the same form factor here in order to regularise the $K^*$ dispersion relation.
  \begin{figure}
    \centering
    \includegraphics[width=0.45\textwidth]{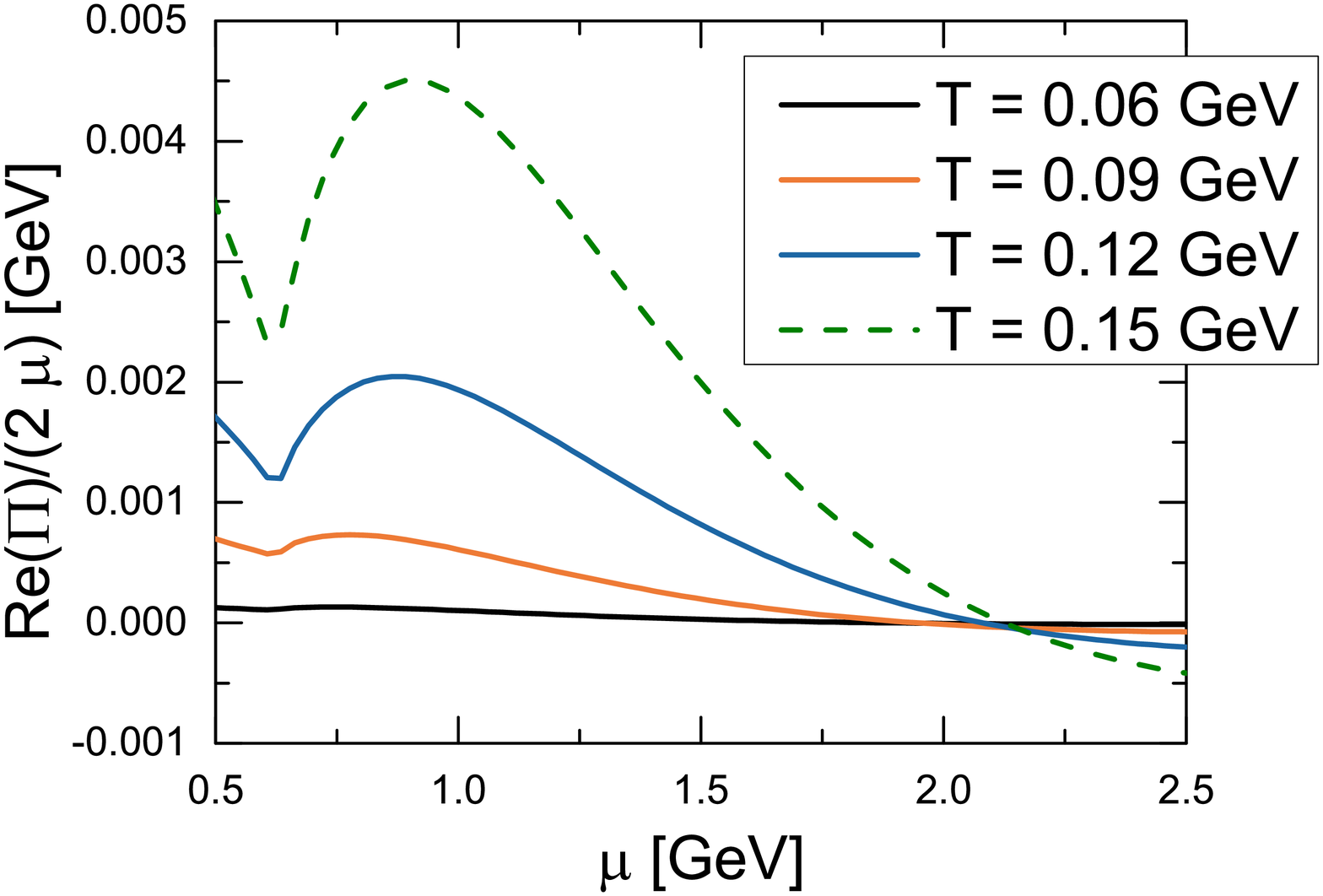}
    \includegraphics[width=0.45\textwidth]{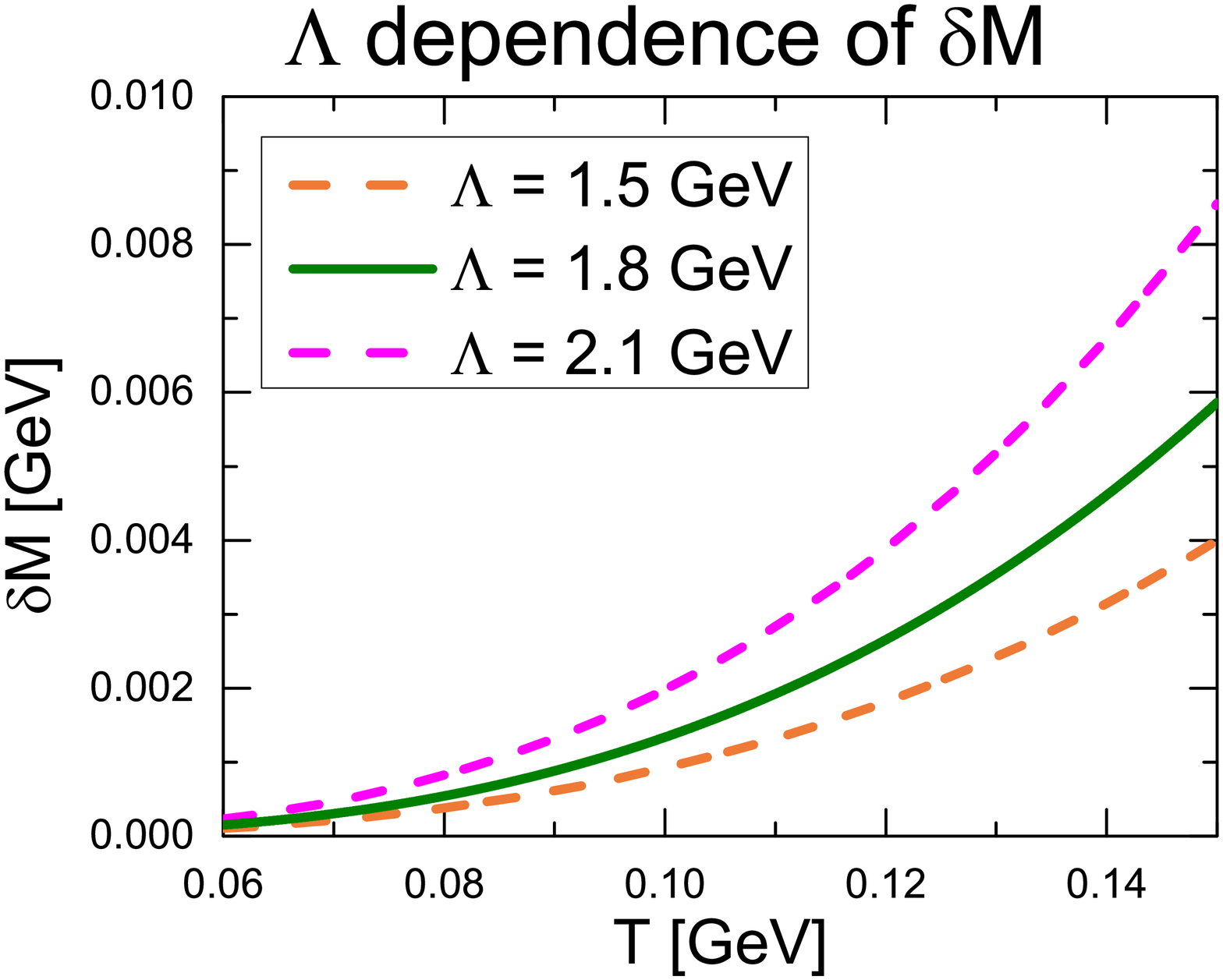}
    \caption{Left: Real part of the $K^*$ selfenergy at different temperatures up to $T=150$~MeV.
    Right: $K^*$ mass shift as a function of the temperature for $\Lambda = 1.5,1.8,2.1$~GeV (lower, middle and upper curves, respectively).}
    \label{fig:tksdm}
  \end{figure}
For our calculation of the $K^*$ properties we bracket the uncertainty due to the unconstrained $\Lambda$ parameter in the range $\Lambda=1.5$-$2.1$~GeV. For instance, for $\Lambda=1.8$~GeV, the dispersion relation has reached convergence at the level of per mil for $\mu\simeq3$~GeV. By inspecting the differential decay distribution of the $K^*$ width [integrand of Eq.~(\ref{eq:Kstar-widths})] one can infer the mean value of the momentum $q$ relevant for the decay process and we find that the form factor is cutting off momenta beyond a typical hadronic scale of 1.5-2~GeV$/c$.

The real part of the $K^*$ selfenergy, as obtained from Eq.~(\ref{eq:DR-reg}), is shown in Fig.~\ref{fig:tksdm} (left panel). It exhibits a sizable energy dependence associated to the quick rise in $\Gamma^*_{K^*}$ beyond the threshold, as expected from a $p$-wave decaying resonance. We note that the absolute size of $\textrm{Re}\, \Pi_K^*$ is below 10~MeV and of repulsive nature, practically negligible up to $T=100$~MeV and the larger values being reached at the highest temperature of $T=150$~MeV. The right panel of Fig.~\ref{fig:tksdm} shows the corresponding $K^*$ mass shift as a function of the temperature for three different values of $\Lambda$.  The spread between the different values of $\Lambda$ becomes larger with increasing temperatures. We note, however, that the mass shift is in itself already much smaller than $M_K^*$, thus rendering the uncertainty due to the form factor unimportant. Nevertheless we find that, although we have changed the form factor parameter within a large range of values as compared to typical hadronic scales, the real part of the selfenergy at the $K^*$ mass stays repulsive (does not flip sign), an indication of stability of the dispersion relation.
  \begin{figure}
    \centering
    \includegraphics[width=0.38\textwidth]{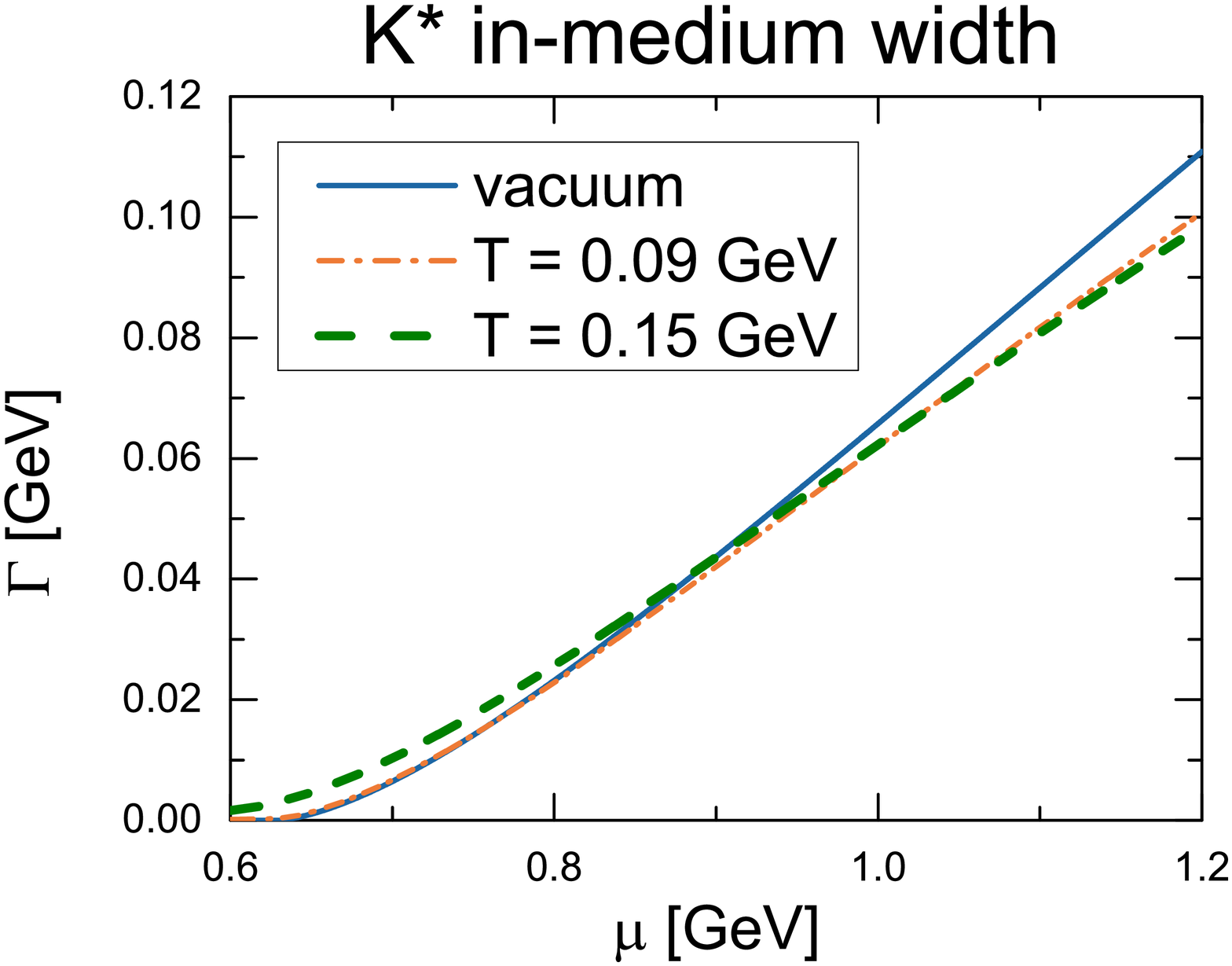}
    \includegraphics[width=0.61\textwidth]{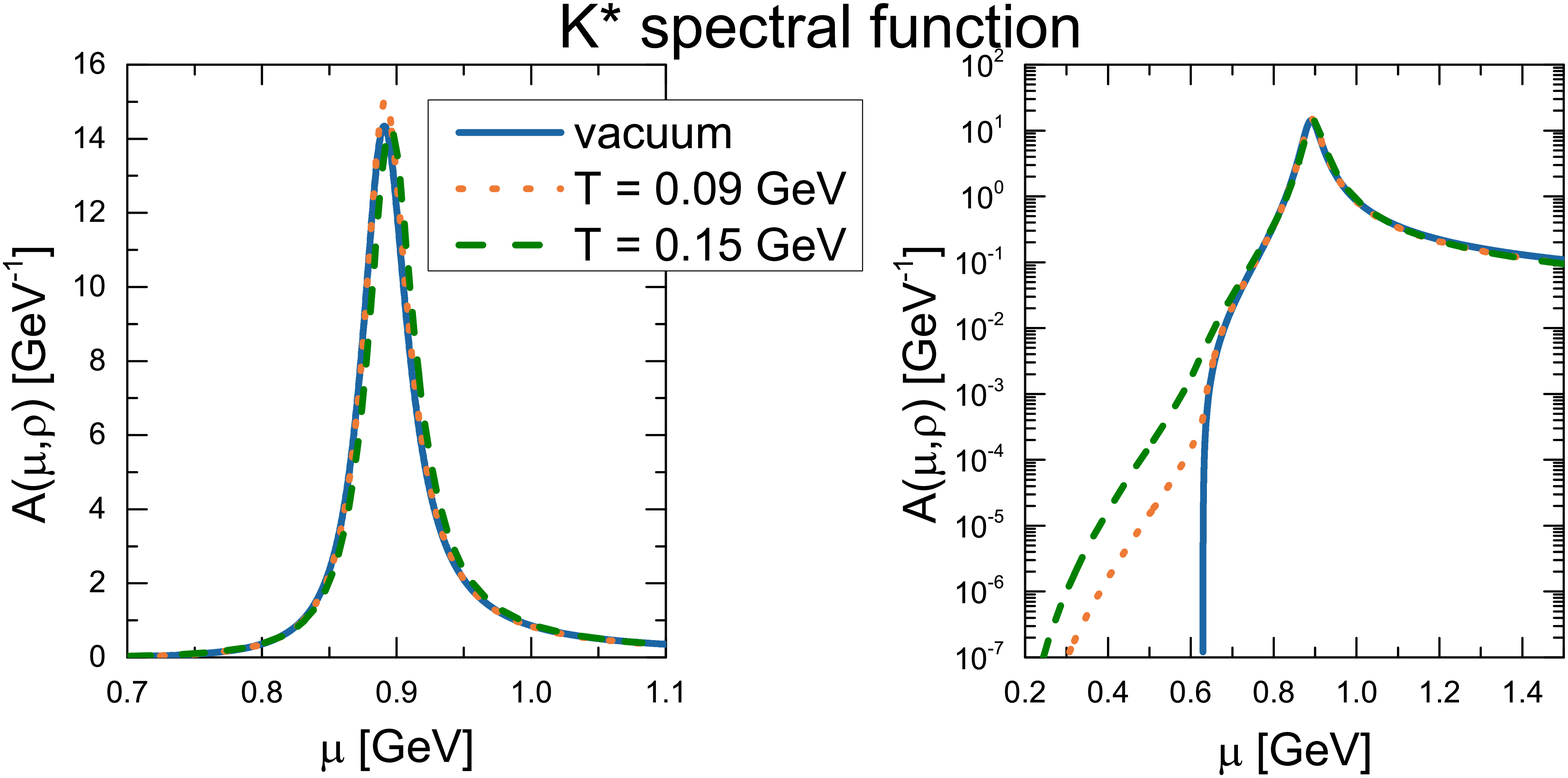}
    \caption{Left: Off-shell in-medium width of the $K^*$ in hot matter as a function of energy at several temperatures. Middle (Right): Spectral function (in log-scale) of the $K^{*}$ including both the mass shift and in-medium width in hot matter for different temperatures.}
    \label{fig:tkssf}
  \end{figure}

In Fig.~\ref{fig:tkssf} (left) we depict the $K^*$ in-medium width as a function of the energy at several temperatures. The width is slightly enhanced in the lower mass region with increasing temperatures due to the attraction experienced by the kaons and their broader spectral function, which augments the density of available $\bar K (K) \pi$ states to decay in. We find that the sensitivity of $\Gamma^*_{K^*}$ to the in-medium properties of the kaons is small, in agreement with the findings of \cite{Tolos:2010fq} in cold nuclear matter, and only at the highest temperatures some broadening is expectable below the $K^*$ mass. In Ref.~\cite{Faessler:2002qb} the results for the kaon selfenergy have been used to estimate a factor 3 increase in the $\phi$ meson decay width close to the critical temperature. Such a large sensitivity of the $\phi$ meson is due to (i) the main decay mode being $\bar K K$ and (ii) the fact that this decay occurs slightly above threshold in vacuum. We also note that we have neglected the effect of Bose enhancement at finite temperature. As it has been discussed in \cite{Cabrera-proc,Tolos:2008di}, for the nominal masses of the daughter particles Bose enhancement is not relevant at the expected temperatures in the hadronic phase of a heavy-ion collision. However, it may become important if $K$ and $\pi$ develop low-energy excitations, which is likely to be the case in hot matter at finite baryonic chemical potential (FAIR conditions) \cite{Cabrera-proc}.

In the middle and right panels of Fig.~\ref{fig:tkssf} one can see the resulting spectral function for the $K^{*}$.  As expected from the discussion above the $K^{*}$ peak experiences a very small shift to higher energies from the interaction with the medium. In spite of the changes in the width, the effect is barely noticeable in the spectral function, since those are moderate and take place in the low-energy region. A logarithmic plot of the spectral function (c.f.~Fig.~\ref{fig:tkssf} right) shows the additional strength below the nominal $K\pi$ threshold.
The present result justifies the description of the $K^*$ in a hot environment as a narrow quasiparticle (with properties close to vacuum ones) in transport approaches.

  \section{Summary}
\label{sec:summary}

We have presented a study of medium effects on the light strange vector mesons in different scenarios of hot and dense strongly interacting matter, namely $(\mu_B\ne 0, T\simeq 0)$ and $(\mu_B\simeq 0, T \ne 0)$.

In the first case, the $K^*$ and $\bar K^*$ mesons behave rather differently, due to the possibility of exciting near-threshold $S=-1$ baryons by the $\bar K^*$, a mechanism that is absent for the $K^*$ due to strangeness conservation.
We have evaluated the collisional part of the $K^*$ selfenergy in a chiral unitary approach based on the hidden local gauge symmetry, which was recently applied to describe the interaction between vector mesons and baryons. Similarly to what has been found for the $K$ meson in previous studies, the $K^*$ experiences a mild repulsive mass shift from the interaction with nucleons, of about 5\% at normal nuclear matter density. The decay width of the $K^*$ is somewhat modified by the properties of the $K$ in the medium (pionic effects at finite density, however, may provide a more sizable enhancement of the $K\pi$ decay mode \cite{Tolos:2010fq,Cabrera-proc}). Overall, the spectral function of the $K^*$ maintains a single-peak structure, which facilitates its treatment in microscopic transport approaches \cite{Hartnack:2011cn}, particularly those incorporating off-shell effects \cite{Cassing:2003vz}.
For the $\bar K^*$ however, the quasiparticle picture within a Breit-Wigner parameterisation of the spectral function is difficult to be reconciled with existing hadronic many-body calculations, which point at a broad, multi-component structure as a consequence of the mixing with several collective excitations involving strange baryon resonances. In this case the Breit-Wigner parameterisation, with in-medium parameters extracted from the $\bar K^*$ selfenergy, can only describe qualitatively the behaviour of the $\bar K^*$ spectral function at finite nuclear density. Whereas the characteristic features of the quasiparticle mode are retained, the details of the collective modes excited in the system as well as the energy dependence inherent to the opening of in-medium channels, are clearly missed.

In the second case (hot hadronic matter with low net baryon content), the major modifications in the $\bar K^* (K^*)$ properties are tied to their predominant decay mode, $\bar K (K) \pi$. Based on a phenomenological estimation of the kaon selfenergy, consistent with Chiral Perturbation Theory at low energies, we have calculated the in-medium vector meson width at finite temperature. The latter increases moderately with temperature at energies below the $K^*$ mass up to the highest considered temperature of 150~MeV (the effect being practically negligible up to $T\lesssim 100$~MeV), due to the kaon experiencing a mildly attractive potential and a moderate broadening from interactions with the (predominantly) pionic gas. The changes in the $K^*$ width are accompanied by a small repulsive mass shift, which we have calculated by means of a dispersion relation. We conclude that the $K^*$ resonances are only mildly sensitive to the hadronic medium in absence of baryons for the typical temperature range expected in the hadronic phase of a heavy-ion collision.

The present study aims to facilitate the analysis of the in-medium properties of strange resonances in heavy-ion collisions within microscopic transport approaches, a topic of on-going experimental activity within the RHIC low-energy scan programme and the HADES Collaboration at GSI. Implementation of in-medium spectral functions in the Breit-Wigner prescription provides a fair description of the underlying many-body dynamics, with the exception of the $S=-1$ sector at dense matter, where only an average, qualitative representation of the features of microscopic models can be achieved. The omission of medium effects on pions is not expected to play a significant role in hot hadronic matter, whereas at $\mu_B\ne 0$ it could be mimicked by an increase of the in-medium decay rate in the case of the $K^*$.

\section*{Acknowledgements}
We acknowledge fruitful discussions with J\"org Aichelin, Wolfgang Cassing, Laura Fabbietti, Christina Markert, Eulogio Oset, \`Angels Ramos and Laura Tol\'os. This work has been supported by the Helmholtz International Center for FAIR within the framework of the LOEWE program. D.C. acknowledges support from the BMBF (Germany) under project no.~05P12RFFCQ and A.I. acknowledges financial support from the HGS-HIRe for FAIR and H-QM.

\end{document}